\title{On a class of third order mappings with two rational invariants}
\author{V.E. Adler}
\date{\empty}
\documentclass[a4paper,12pt]{article}
\usepackage{amsmath,amssymb,amsthm,epic,eepic,graphicx}
\usepackage[english]{babel}
\usepackage[usenames,dvipsnames]{color}
\usepackage{hyperref}

\def\ti{\tilde}
\def\id{\mathop{\rm id}}

\def\xyz{\begin{pmatrix}x\\ y\\ z\end{pmatrix}}

\newtheorem{theorem}{Theorem}
\newtheorem{ex}{Example}
\newenvironment{example}{\begin{ex}\rm}{\end{ex}}

\begin{document}
\maketitle
\thispagestyle{empty}
\def\figurename{Fig.}

\let\oldf=\thefootnote \renewcommand{\thefootnote}{}
\footnote{\hspace{-7mm} L.D.~Landau Institute for Theoretical Physics, RAS\\
 1-A pr.~Ak.~Semenova, 142432 Chernogolovka, Russia\\
 E-mail: {\tt adler@itp.ac.ru}\\
 The work was supported by the RFBR grant 04-01-00403}
\let\thefootnote=\oldf

\begin{abstract}
A novel family of integrable third order maps is presented. Each map
possesses, by construction, a pair of rational invariants and a commuting
map from the same class. The 3-dimensional invariant curve is parametrized,
in general, by an elliptic curve.
\end{abstract}

%-------------------------------------------------------------------------------
\section{Introduction}

The construction of many nontrivial rational mappings is based on the
trivial idea: if one root of a quadratic equation is known then the second
one is found in rational form. The most important example is the family of
QRT mappings \cite{QRT} introduced as follows: let $f(x,y)$, $g(x,y)$ be
biquadratic polynomials and $I=f/g$, then the corresponding map
$(x,y)\to(\ti x,\ti y)$ is defined by equations
\[
 I(x,y)=I(\ti x,y)=I(\ti x,\ti y)
\]
where the solutions $\ti x=x$, $\ti y=y$ are ignored. Obviously, the
resulting map is the composition of two rational involutions and $I$ is its
invariant by construction.

This can be generalized in several ways. For example, one may consider the
rational mapping $(x,y,z)\to(\ti x,\ti y,\ti z)$ defined by equations
\[
 I(x,y,z)=I(\ti x,y,z)=I(\ti x,\ti y,z)=I(\ti x,\ti y,\ti z)
\]
where $I=f/g$ is ratio of two three-quadratic polynomials. The
generalization for any number of variables is straightforward. In contrast
to the QRT case such mappings are not integrable in general. However, in
papers \cite{Iatrou03a,Iatrou03b} some instances were found when a second
polynomial or rational invariant exists. Several other examples of third
order integrable mappings were studied in
\cite{Hirota_Kimura_Yahagi,Matsukidaira_Takahashi}.

In this paper we consider another possibility, assuming from the beginning
that the map under construction possesses two invariants $I,J$, but on each
step two of three variables are changed, rather than one. The idea of
differencing two invariants simultaneously was used previously in paper
\cite{Roberts_Iatrou_Quispel}, but in a different manner. It is easy to see
that if $I$ and $J$ are ratios of affine-linear polynomials on $x,y,z$ (that
is, numerators and denominators are of the first degree on each argument)
then the resulting map is a composition of rational involutions again. Three
basic involutions satisfy a simple identity and as a result one obtains two
commuting maps which share the common invariants and generate the kagome
lattice. Recall that the existence of commuting partner is a typical feature
of integrable maps \cite{Veselov91}. The generalization for any number of
variables is also straightforward, however the resulting maps are not in
general integrable.

It should be stressed that the presented family in no way covers all
integrable cases of third order maps. Actually, it seems to be not so thick
as the QRT-like maps from the papers cited above. On the other hand, the
construction scheme seems to be more explicit. Of course, there should be
intersections between these families, but the absence of any classification
results makes the detailed comparison impossible for the present. Anyway, we
see again that the most trivial ideas work more than once (cf. also the
recent construction of Yang-Baxter maps \cite{ABS}).

%-------------------------------------------------------------------------------
\section{The triad mapping}

Let $f,g,h,k$ be affine-linear polynomials on $x,y,z$ and $I=f/g$, $J=h/k$.
Consider equations for the unknowns $\ti y,\ti z$:
\[
 I(x,y,z)=I(x,\ti y,\ti z),\quad J(x,y,z)=J(x,\ti y,\ti z).
\]
This is a system of the form
\[
 a\ti y\ti z+b\ti y+c\ti z+d=0,\quad A\ti y\ti z+B\ti y+C\ti z+D=0
\]
with coefficients depending rationally on $x,y,z$. Obviously, it is
equivalent to a quadratic equation and since the solution $(\ti y,\ti
z)=(y,z)$ is known, hence the second solution can be easily found in
rational form. This defines the map $(x,y,z)\mapsto(x,\ti y,\ti z)$.
Changing the roles of the variables we obtain three rational mappings
defined by equations
\begin{align*}
 R_1:&\quad I(x,y,z)=I(x,y_1,z_1),\quad J(x,y,z)=J(x,y_1,z_1),\\
 R_2:&\quad I(x,y,z)=I(x_2,y,z_2),\quad J(x,y,z)=J(x_2,y,z_2),\\
 R_3:&\quad I(x,y,z)=I(x_3,y_3,z),\quad J(x,y,z)=J(x_3,y_3,z)
\end{align*}
assuming that the identical solutions are always ignored. By construction,
these maps are involutive:
\begin{equation}\label{Rid1}
 R^2_1=R^2_2=R^2_3=\id.
\end{equation}
The following property is far from being obvious.

%------------------------------------------------------------
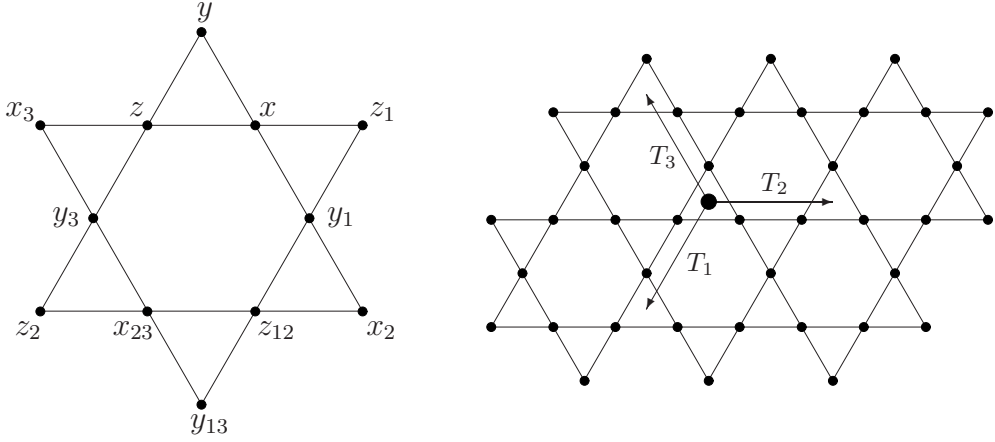
\begin{figure}[tb]
\begin{center}
\setlength{\unitlength}{0.03em}
\def\cix{\color{red}\circle*{10}}
\def\ciy{\color{green}\circle*{10}}
\def\ciz{\color{blue}\circle*{10}}
%--------------------------------------------
\begin{picture}(400,450)(-200,-250)
 \path(173, 100)(-173, 100)(0,-200)(173,100)
 \path(173,-100)(-173,-100)(0, 200)(173,-100)
 \put(173, 100){\ciz} \put(-173, 100){\cix} \put(0,-200){\ciy}
 \put(173,-100){\cix} \put(-173,-100){\ciz} \put(0, 200){\ciy}
 \put( -58, 100){\ciz} \put( 58, 100){\cix}
 \put(-116,   0){\ciy} \put(116,   0){\ciy}
 \put( -58,-100){\cix} \put( 58,-100){\ciz}
 \put(-5,218){$y$}
 \put(-210,110){$x_3$} \put(-78,110){$z$}
   \put(62,110){$x$} \put(180,110){$z_1$}
 \put(-160,-3){$y_3$} \put(135,-3){$y_1$}
 \put(-200,-125){$z_2$} \put(-95,-125){$x_{23}$}
   \put(60,-125){$z_{12}$} \put(177,-125){$x_2$}
 \put(-12,-225){$y_{13}$}
\end{picture}\qquad
%--------------------------------------------
\setlength{\unitlength}{0.01em}
\def\cix{\color{red}\circle*{30}}
\def\ciy{\color{green}\circle*{30}}
\def\ciz{\color{blue}\circle*{30}}
%--------------------------------------------
\begin{picture}(1600,1300)(0,-400)
 \path(0,0)(500,865)(1100,-173)(1600,692)(200,692)(700,-173)
      (1300,865)(1600,346)(0,346)(300,-173)(900,865)(1400,0)(0,0)
 \path(700,404)(500,58) \put(510, 78){\vector(-1,-2){10}}
   \put(630,180){\footnotesize$T_1$}
 \path(700,404)(500,750)\put(510,730){\vector(-1, 2){10}}
 \put(700,404){\vector(1,0){400}}
   \put(870,440){\footnotesize$T_2$}
 \put(700,404){\color{white}\circle*{50}}
    \put(510,525){\footnotesize$T_3$}
\multiput(200,  0)(400,0){4}{\cix}
 \multiput(  0,346)(400,0){5}{\cix}
 \multiput(200,692)(400,0){4}{\cix}
     \multiput(300,-173)(400,0){3}{\ciy}
     \multiput(100, 173)(400,0){4}{\ciy}
     \multiput(300, 519)(400,0){4}{\ciy}
     \multiput(500, 865)(400,0){3}{\ciy}
 \multiput(0,0)(200,346){3}{\multiput(0,0)(400,0){4}{\ciz}}
 \end{picture}
%--------------------------------------------
\caption{Graphical representation of the identities (\ref{Rid1}), (\ref{Rid2})
and the kagome lattice}
\label{fig:kagome}
\end{center}
\end{figure}
%------------------------------------------------------------

\begin{theorem}
Maps $R_i$ satisfy the identity
\begin{equation}\label{Rid2}
  R_1R_2R_3=R_3R_2R_1.
\end{equation}
\end{theorem}
\begin{proof}
We present the computational proof based on the fact that all points lie on
the invariant curve $f=Ig$, $h=Jk$ which is the intersection of two surfaces
of the form
\begin{equation}\label{affine}
 A:\quad a_1XYZ+a_2XY+a_3XZ+a_4YZ+a_5X+a_6Y+a_7Z+a_8=0.
\end{equation}
Consider five points
\begin{equation}\label{points}
 (x_2,y_1,z_{12})\stackrel{R_2}{\leftarrow}(x,y_1,z_1)
 \stackrel{R_1}{\leftarrow}(x,y,z)\stackrel{R_3}{\to}(x_3,y_3,z)
 \stackrel{R_2}{\to}(x_{23},y_3,z_2)
\end{equation}
(the enumeration is shown on the fig.~\ref{fig:kagome}). It is sufficient to
prove that the values of $y_{13}$ obtained in two different ways coincide.
This is equivalent to the statement that the invariant curve intersects the
straight line $L:$ $(X,Z)=(x_{23},z_{12})$.

Let $L\cap A=(x_{23},y^*,z_{12})$. Consider the intersection lines of the
surface (\ref{affine}) with the planes $X=x$, $Y=y_3$ and $Z=z_{12}$. Let
$(\xi,y_3,z_{12})$, $(x,\eta,z_{12})$ and $(x,y_3,\zeta)$ be the mutually
common points of these lines. Then the following equations holds:
\begin{alignat*}{3}
 &X=x &&:\quad&& \det([y,z],[y_1,z_1],[\eta,z_{12}],[y_3,\zeta])=0,   \\
 &Y=y_3    &&:&& \det([x_3,z],[x_{23},z_2],[x,\zeta],[\xi,z_{12}])=0,\\
 &Z=z_{12} &&:&& \det([x_2,y_1],[x_{23},y^*],[\xi,y_3],[x,\eta])=0
\end{alignat*}
where the notation $[p,q]=(pq,p,q,1)^T$ is used.

Now we find $y^*$ from the last equation where $\xi$ and $\eta$ are
eliminated by use of the first and second ones. The remarkable fact, proved
by direct and rather tedious computation is that $\zeta$ cancels out and
$y^*$ does not depend on it. This means that any surface of the form
(\ref{affine}) passing through five points (\ref{points}) passes also
through the point $(x_{23},y^*,z_{12})$. Since the invariant curve is the
intersection of two such surfaces, it also runs through this point and
$y_{13}=y^*$.
\end{proof}

As a corollary we immediately obtain that the mappings
\[
 T_1=R_2R_3,\quad T_2=R_3R_1,\quad T_3=R_1R_2
\]
satisfy the identities
\begin{equation*}\label{Tid}
 T_iT_j=T_jT_i,\quad T_1T_2T_3=\id.
\end{equation*}
Therefore we have, in general, a pair of commuting mappings which generate
the kagome lattice. However, for some special choices of $I,J$ this lattice
may be reduced due to additional identities for the generators, see Examples
\ref{ex:reduced}, \ref{ex:finite}.

The projection of the invariant curve onto the coordinate plane $(x,y)$ is
defined by the equation $b(x,y)=F\partial_zH-H\partial_zF=0$ where $F=f-Ig$,
$H=h-Jk$. Generically, this is a biquadratic curve of genus 1, and the
invariant curve is parametrized by the point on the elliptic curve
$X^2=r(x)=(\partial_yb)^2-2b\partial^2_yb$.

%-------------------------------------------------------------------------------
\section{Examples}

Here we consider only few very particular examples of the presented
mappings. The investigation of the whole family is probably an interesting
but also a very difficult problem. The size of this family can be estimated
roughly as follows. An affine-linear polynomial on $x,y,z$ contains $2^3$
coefficient parameters. In the ratios $I,J$ two parameters are scaled out
and we also have to take into account the 3-parametric group of M\"obius
transformations which acts on each variable independently as well as on the
invariants. Therefore, the total number of essential parameters in the
mappings under consideration is at most $4\cdot2^3-2-(2+3)\cdot3=15$ (the
analogous reasoning gives $2\cdot3^2-1-(1+2)\cdot3=8$ for the QRT mappings
and $2\cdot3^3-1-(1+3)\cdot3=41$ for its 3-component generalization
mentioned in Introduction).

%-------------------------------------------------------------------------------
\begin{example}\label{ex:generic}{\bf A generic mapping.}
The simplest way to get some experience is to generate maps for the random
choice of the coefficeints in $I,J$ and to iterate the random initial data.
It turns out that already coefficients with the random values $0,\pm1$
provide, as a rule, the nondegerate map. The fig.~\ref{fig:generic} plots
the images of the point $(x,y,z)=(1/2,1/2,-1)$ under the mappings $T_1$ and
$T_2$ for the invariants
\[
 I=\frac{y+z+xy-xyz}{x-z+xy},\quad
 J=\frac{1+x-z-xy-xz-yz-xyz}{1-x+y-z-xy-xz+yz-xyz}.
\]
Here we see that $T_2$ runs through only one branch of the invariant curve.
Although the invariants look not too complicated, the corresponding maps are
extremely bulky. For example, three components of $R_1$ contains in total 84
terms and $T_1$ contains 831 terms.
%-----------------------------------------------------------------
\begin{figure}[t]
\centerline{\includegraphics[width=60mm]{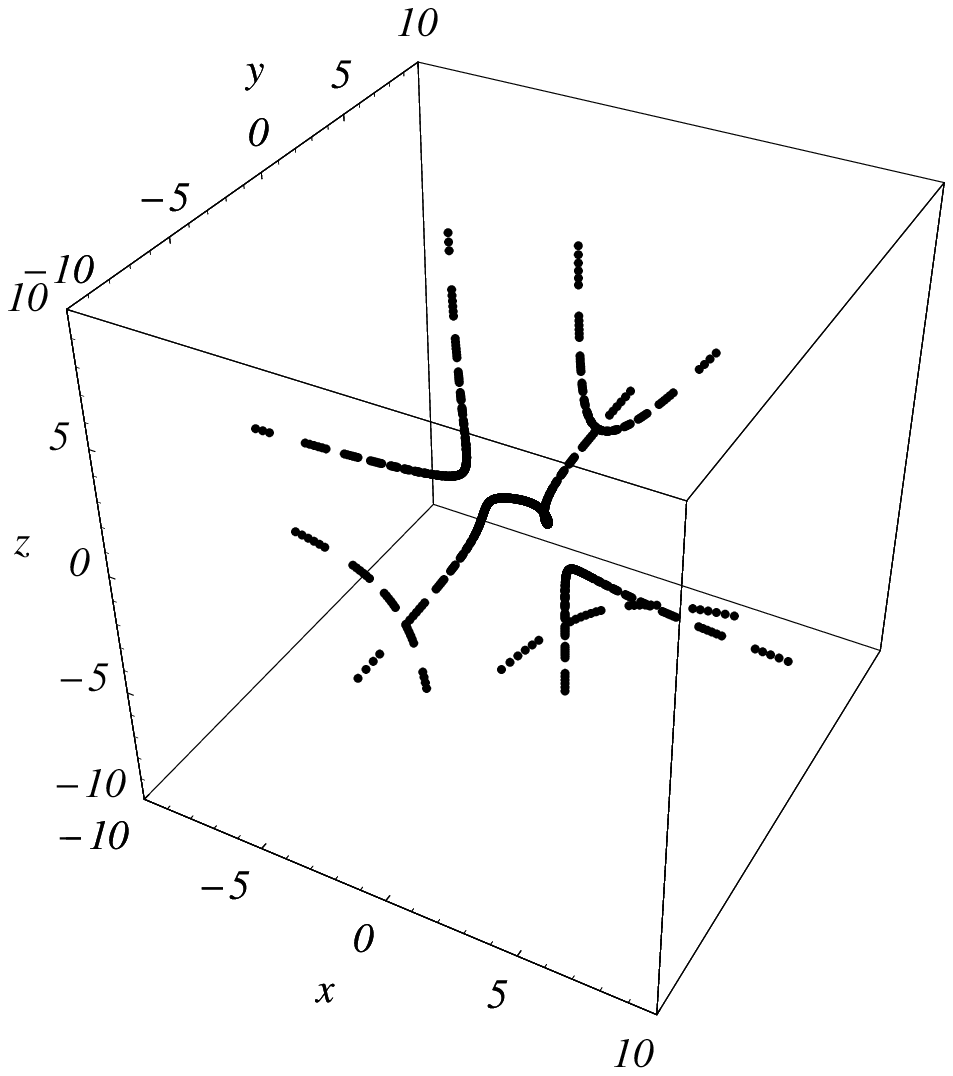}\quad
            \includegraphics[width=60mm]{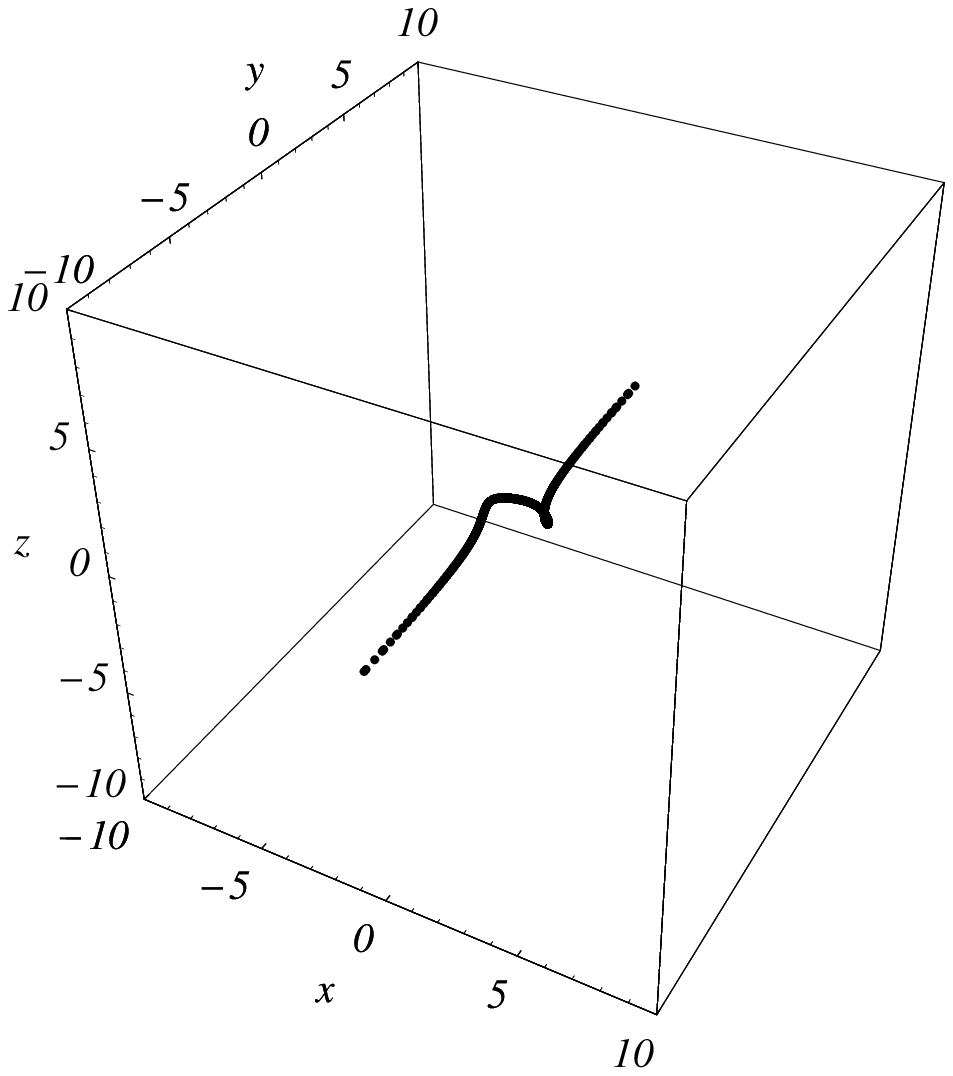}}
\caption{Ex.~\ref{ex:generic}.
 The iterations of $T_1$ and $T_2$.}
\label{fig:generic}
\end{figure}
%-----------------------------------------------------------------
\end{example}

%-------------------------------------------------------------------------------
\begin{example}\label{ex:polynomial}
{\bf A mapping with polynomial invariants.} As in QRT case, more simple, but
still nontrival maps can be obtained already for polynomial invariants. Let
\[
 I=x+y+z-xyz,\quad J=z(x+ay)+bx+cy
\]
then the corresponding involutions are
\begin{gather*}
 R_1\xyz= \begin{pmatrix} x \\
            \dfrac{x^2z-x+az+c}{cx+a} \\
            \dfrac{cxy+x+ay-c}{x^2+a} \end{pmatrix},\quad
 R_2\xyz= \begin{pmatrix}
             \dfrac{ay^2z-ay+z+b}{by+1} \\ y \\
             \dfrac{bxy+x+ay-b}{ay^2+1} \end{pmatrix},\\
 R_3\xyz= \begin{pmatrix}
             \dfrac{ayz^2+cyz+(1-a)z+b-c}{z(z+b)} \\
             \dfrac{xz^2+bxz+(a-1)z+c-b}{z(az+c)} \\ z \end{pmatrix}
\end{gather*}
The iterations of the map $T_1=R_2R_3$ are shown on the
fig.~\ref{fig:polynomial}.
%-----------------------------------------------------------------
\begin{figure}[t]
\centerline{\includegraphics[width=80mm]{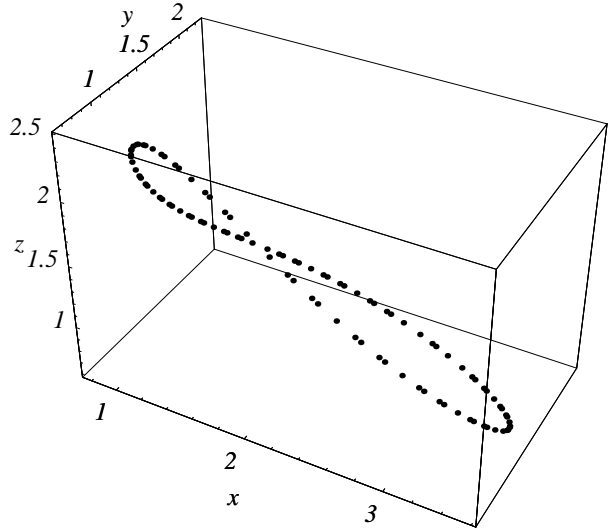}}
\caption{Ex.~\ref{ex:polynomial}.
 Iterations of the mapping $T_1$, $a=3,$ $b=1$, $c=2$;
 initial values $(x,y,z)=(1,2,1)$}
\label{fig:polynomial}
\end{figure}
%-----------------------------------------------------------------
\end{example}

%-------------------------------------------------------------------------------
\begin{example}\label{ex:reduced}{\bf Reduced group.}
In some cases the involutions $R_i$ may satisfy additional identities. For
example, this happens if invariants are symmetric with respect to a pair of
variables. Consider the invariants:
\[
 I=xy+z,\quad J=(x+y)z.
\]
One can check straightforwardly that the corresponding maps
\[
 R_1\xyz= \begin{pmatrix} x \\ z/x-x \\ x(x+y) \end{pmatrix},\quad
 R_2\xyz= \begin{pmatrix} z/y-y \\ y \\ y(x+y) \end{pmatrix},\quad
 R_3\xyz= \begin{pmatrix} y \\ x \\ z \end{pmatrix}
\]
satisfy, in addition to the identities (\ref{Rid1}), (\ref{Rid2}), the
relation $R_2R_3=R_3R_1$. This means that in this case we obtain only one
mapping $T_1=T_2$ while $T_3=T^{-2}_1$. It should be noted that such sort of
group reduction is not related to the polynomiality of the invariants or to
the degeneration of the invariant curve. Indeed, in this example its
projection on $(x,y)$ plane is the curve $(xy-I)(x+y)+J=0$ of genus $1$ iff
$J\ne0$.
%-----------------------------------------------------------------
\begin{figure}[t]
\centerline{\includegraphics[width=60mm]{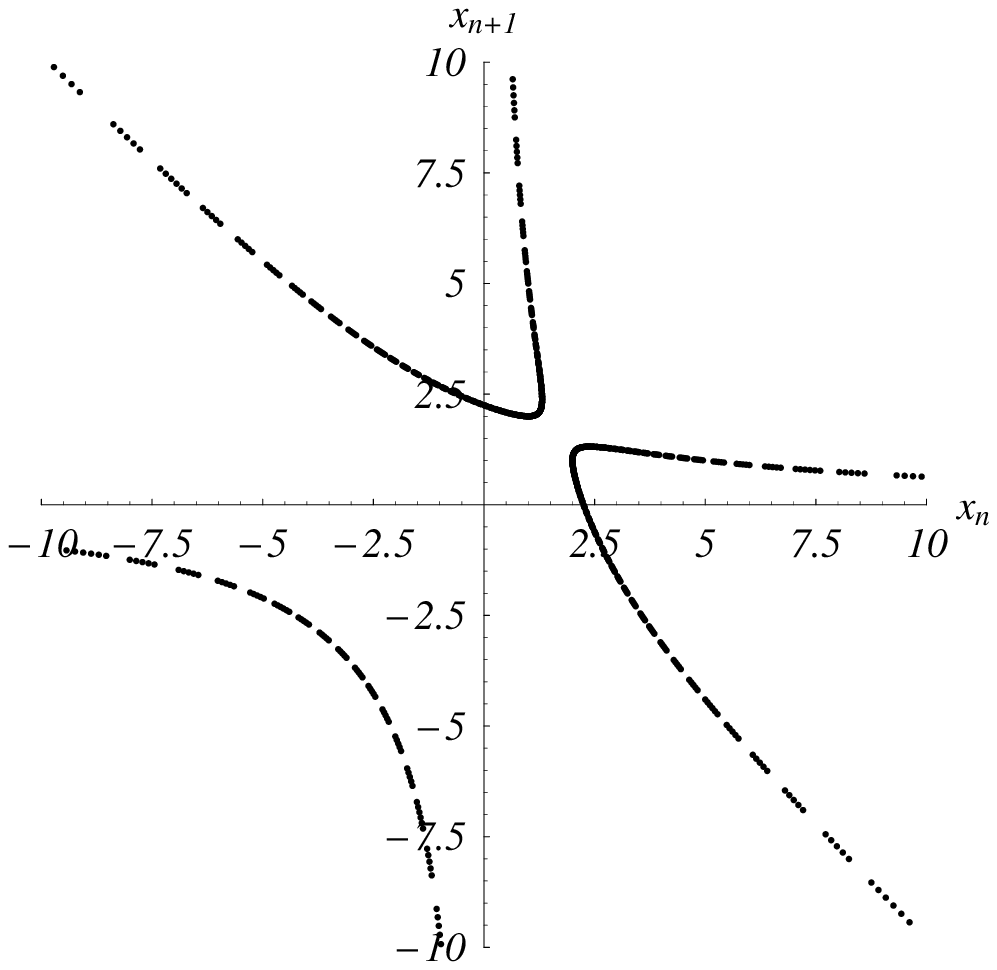}\quad
            \includegraphics[width=60mm]{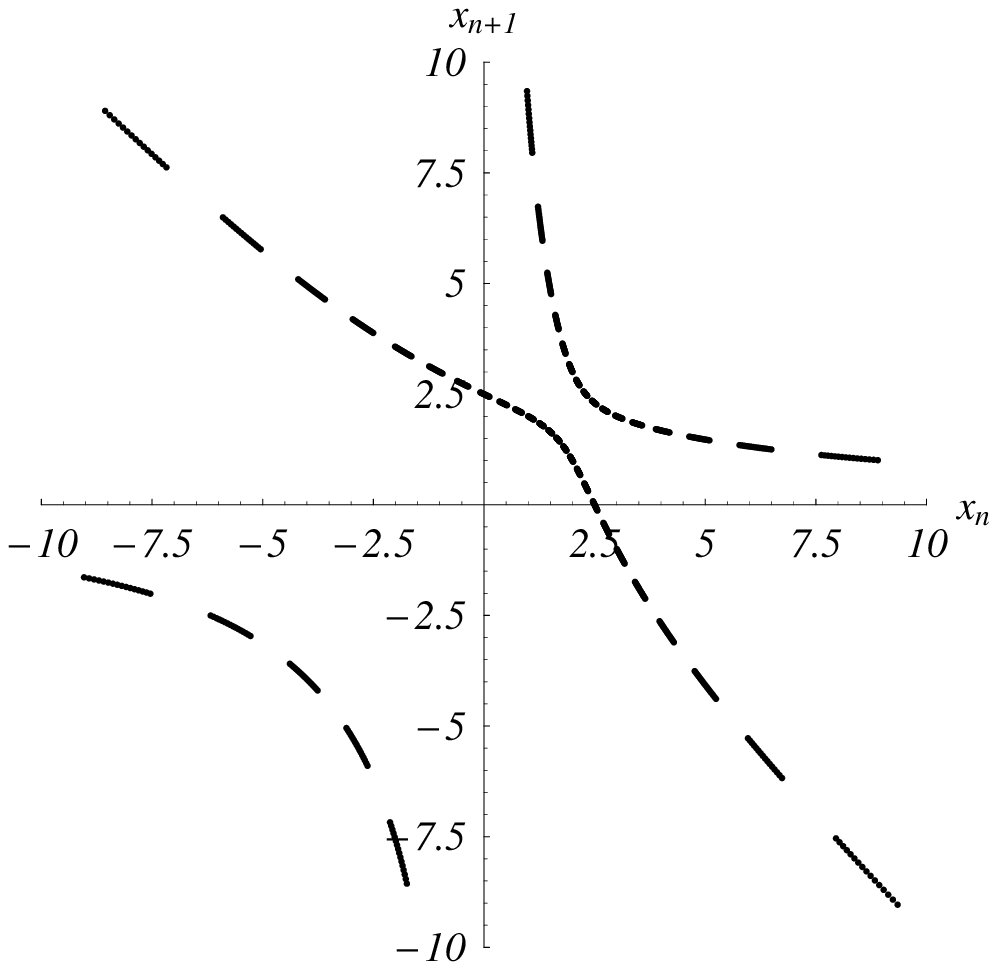}}
\caption{Ex.~\ref{ex:reduced}.
 The iterations of the various initial data}
\label{fig:reduced}
\end{figure}
%-----------------------------------------------------------------

The map $T_1$ generates the discrete system
\[
 \xyz_{n+1}=\begin{pmatrix} z/x-x \\ x \\ x(x+y) \end{pmatrix}_n
\]
which can be easily rewritten as the third-order difference equation (see
fig.~\ref{fig:reduced})
\[
 (x_{n+3}+x_{n+2})x_{n+2}=x_{n+1}(x_{n+1}+x_n).
\]
In this notation the invariants take the form
\[
 I=x_n(x_{n+1}+x_n+x_{n-1}),\quad J=(x_{n+1}+x_n)x_n(x_n+x_{n-1}).
\]
\end{example}

%-------------------------------------------------------------------------------
\begin{example}\label{ex:finite}{\bf Finite group.}
Even more degeneracy occurs when all involutions $R_i$ commute and generate
only a few points on the invariant curve. This happens, for example, if
invariants are symmetric with respect to all variables. Obviously, in this
case $R_i$ are just permutations. Not so trivial example is given by
invariants
\[
 I=xy+z,\quad J=\frac{yz+x}{xz+y}.
\]
Here the involutions
\begin{gather*}
 R_1\xyz= \begin{pmatrix} x \\
             \dfrac{xyz+x^2+z^2-1}{xz+y} \\
             \dfrac{x-x^3+xy^2+yz}{xz+y} \end{pmatrix},\quad
 R_2\xyz= \begin{pmatrix}
             \dfrac{xyz+y^2+z^2-1}{yz+x} \\ y \\
             \dfrac{y-y^3+yx^2+xz}{yz+x} \end{pmatrix},\\
 R_3\xyz= \begin{pmatrix} -x \\ -y \\ z \end{pmatrix}
\end{gather*}
also turn out to be commutative.
\end{example}

%-------------------------------------------------------------------------------
\begin{example}\label{ex:noR1}{\bf One involution is identical.}
Consider the invariants
\[
 I=xy+z,\quad J=\frac{xz}{y+z}.
\]
It is easy to check that the system for the map $R_1$ is equivalent to
\[
 y_1z=yz_1,\quad x(y_1-y)=z-z_1
\]
and has only identical solution. Therefore in this case $R_1$ is actually
absent. However, the rest involtutions still generate the nontrivial mapping
\[
 T_1\xyz=\begin{pmatrix}\dfrac{z(x-1)}{y+z} \\ -y-z \\ x(y+z)\end{pmatrix}.
\]
Its iterations are shown on the fig.~\ref{fig:noR1}. Note that the invariant
curve is rational:
\[
  y=\frac{I(x-J)}{x^2-Jx+J},\quad z=\frac{IJ}{x^2-Jx+J}.
\]
%-----------------------------------------------------------------
\begin{figure}[t]
\centerline{\includegraphics[width=80mm]{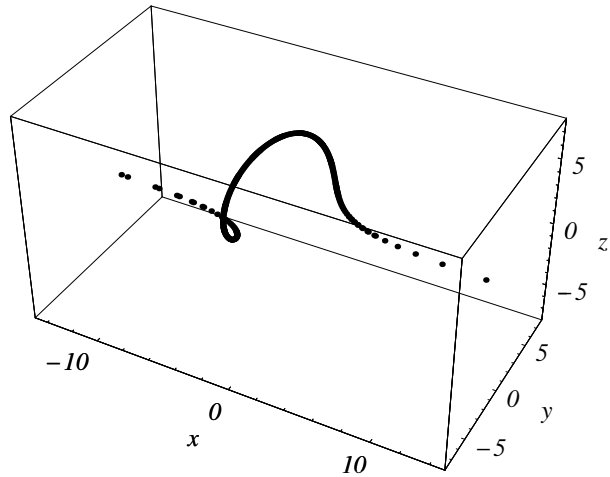}}
\caption{Ex.~\ref{ex:noR1}. The mapping $T_1$.}
\label{fig:noR1}
\end{figure}
%-----------------------------------------------------------------
\end{example}

%-------------------------------------------------------------------------------


\begin{thebibliography}{99}

\bibitem{QRT} G.R.W.~Quispel, J.A.G.~Roberts, C.J.~Thompson.
 Integrable mappings and soliton equations.
 {\em Phys. Lett. A \bfseries 126} (1988) 419--421;
 {\em Physica D \bfseries 34} (1989) 183--192.

\bibitem{Iatrou03a} A.~Iatrou.
 Three dimensional integrable mappings, nlin.SI/0306052.

\bibitem{Iatrou03b} A.~Iatrou.
 Higher dimensional integrable mappings.
 {\em Physica D \bfseries 179} (2003) 229--254.

\bibitem{Hirota_Kimura_Yahagi} R.~Hirota, K.~Kimura, H.~Yahagi.
 How to find the conserved quantities of nonlinear discrete equations.
 {\em J. Phys. A \bfseries 34} (2001) 10377--10386.

\bibitem{Matsukidaira_Takahashi} J.~Matsukidaira, D.~Takahashi.
 Third-order integrable difference equations generated by a pair of
 second-order equations.
 {\em J. Phys. A \bfseries 39} (2006) 1151--1161.

\bibitem{Roberts_Iatrou_Quispel} J.A.G.~Roberts, A.~Iatrou, G.R.W.~Quispel.
 Interchanging parameters and integrals in dynamical systems: the mapping
 case.
 {\em J. Phys. A \bfseries 35} (2002) 2309--2325.

\bibitem{Veselov91} A.P.~Veselov. Integrable maps.
 {\em Russ. Math. Surveys \bfseries 46} (1991) 1--51.

\bibitem{ABS} V.E.~Adler, A.I.~Bobenko, Yu.B.~Suris.
 Geometry of Yang-Baxter maps: pencils of conics and quadrirational
 mappings. {\em Comm. Anal. and Geom. \bfseries 12:5} (2004) 967--1007.


\end{thebibliography}
\end{document}